\documentclass[aps,prl,twocolumn,showpacs,longbibliography,superscriptaddress]{revtex4-1}

\usepackage{color}

\usepackage{graphicx}

\begin{document}
\newcommand{\ud}{{\mathrm d}}
\newcommand{\sech}{\mathrm{sech}}

\title{Hidden symmetries, instabilities, and  current suppression in Brownian ratchets}

\author{David Cubero}
\email[]{dcubero@us.es}
\affiliation{Departamento de F\'{\i}sica Aplicada I, EUP, Universidad de Sevilla, Calle Virgen de \'Africa 7, 41011 Sevilla, Spain}
\author{Ferruccio Renzoni}
\email[]{f.renzoni@ucl.ac.uk}
\affiliation{Department of Physics and Astronomy, University College London, Gower Street, London WC1E 6BT, United Kingdom}

\begin{abstract}
 The operation of Brownian motors is usually described in terms of out-of-equilibrium  and symmetry-breaking settings, with the relevant spatiotemporal symmetries identified from the analysis of the equations of motion for the system at hand. When the appropriate conditions are satisfied, symmetry-related trajectories with opposite current are thought to balance each other, yielding suppression of transport. The direction of the current can be precisely controlled around these symmetry points by finely tuning the driving parameters. 
Here we demonstrate, by studying a prototypical Brownian ratchet system, the existence of  {\it  hidden} symmetries, which escape the identification by the standard symmetry analysis, and require different theoretical tools for their revelation. Furthermore,  we show that system instabilities may lead to  spontaneous symmetry breaking  with unexpected generation of directed transport. 
\end{abstract}

\maketitle

Motion at the nanoscale presents features very different from those encountered in the macroscopic world. Noise is a dominant process at such a scale, and may contribute constructively to the dynamics rather than play the usual role of a disturbance. New mechanisms of transport emerge at the nanoscale, and in particular directed motion may occur  in the absence of an applied bias force. Brownian ratchets \cite{astumian97,reimann02,hanmar09}, the archetypal model system capturing the mechanisms behind such a transport process, represent a key to understand several 
biological processes \cite{mahadevan,schliwa2003}, as well as they inspired a plethora of new  nanodevices displaying directed motion
 \cite{leigh04,valenzuela2010,rou94,bader99,chou99,vanou99,mul03,villegas2003,silva2006,linke1999,salger2009,drexler2013,leigh07,kel99,siegel05,rou94,gommers2005}.  
All these systems are usually described in  terms of operation away from thermal equilibrium, with directed motion following from the breaking of certain spatiotemporal symmetries, identified from the analysis of the equations of motion for the system at hand. Here we prove the existence of  {\it  hidden} symmetries, which escape the identification by the standard symmetry analysis \cite{flach2000,reimann02,hanmar09,denisov14}, and require different theoretical tools for their revelation. 
The main assumption of the standard symmetry analysis, i.e. that two trajectories that are connected by a symmetry transformation carry the same statistical weight, a reasoning which can be traced back to Loschmidt's paradox \cite{loschmidt1876}, yields incorrect predictions in these dissipative systems, failing to account for system instabilities that lead to spontaneous symmetry breaking.

\paragraph{Results ---}
\begin{figure}
\includegraphics[width=8cm]{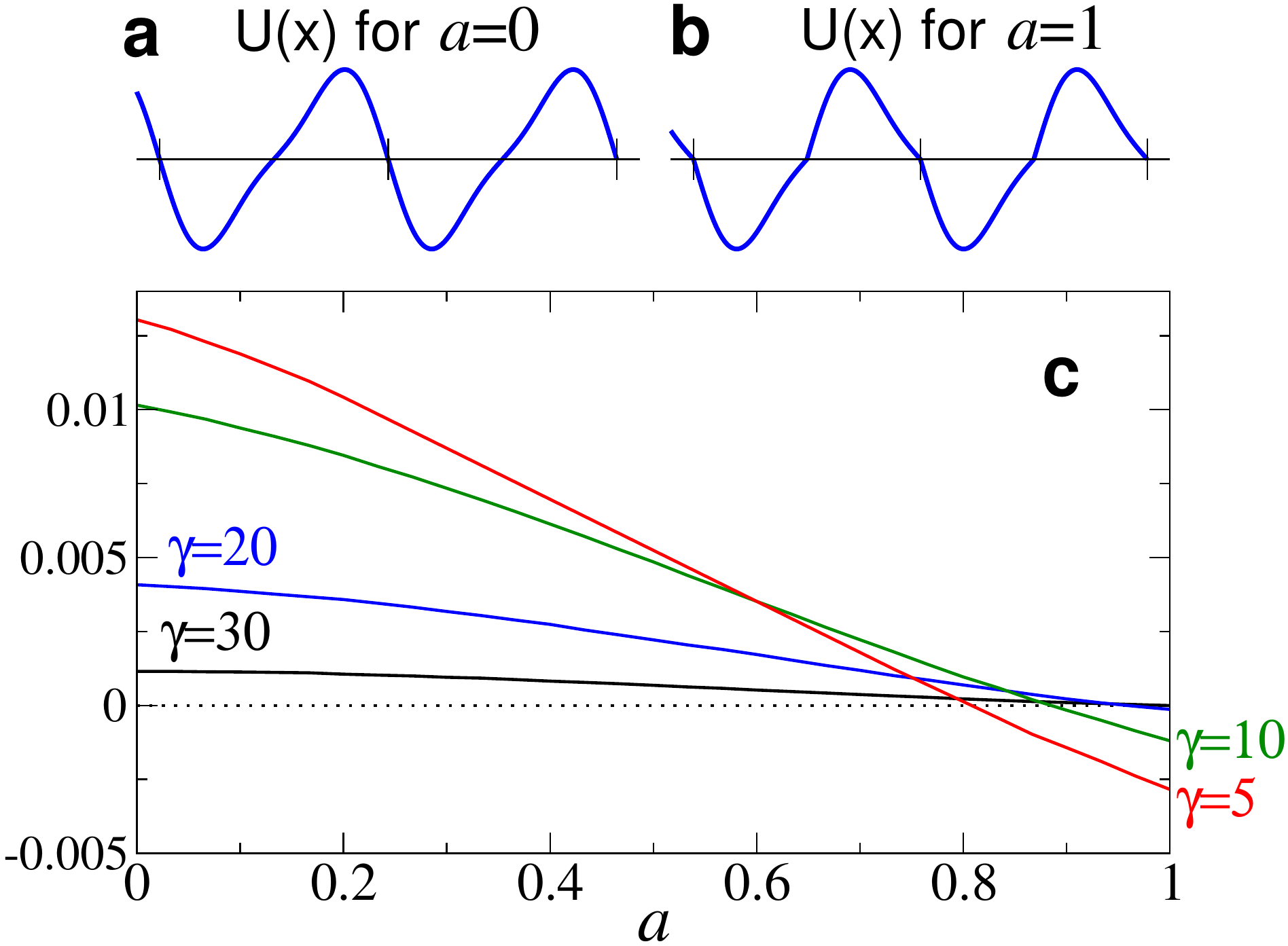}
\caption{
Shift-symmetric potentials act like spatially symmetric ones in one-dimensional overdamped systems.
a) Ratchet potential  $U_\mathrm{rat}(x)=-U_0[\sin(k x)+(1/4)\sin(2kx)]$ with period  $L=2\pi/k$. b) Shift-symmetric potential  defined from  $U_\mathrm{rat}(x)$ as $U_\mathrm{ss}(x)=U_\mathrm{rat}(x)$ in the first half-period, and $U_\mathrm{ss}(x)=-U_\mathrm{ss}(x-L/2)$ in the second half-period.
c) Directed current for a Brownian particle subject to the mixed potential $U(x)=U_\mathrm{rat}(x)(1-a)+U_\mathrm{ss}(x)a$ and  to a shift-symmetric force defined by $F(t)= g(t)\equiv A[\sin(\omega t)+(1/4)\sin(2\omega t)]$ in the first half-period, and $F(t)=-g(t-\tau/2)$ in the second half, where $\tau=2\pi/\omega$. Reduced units are defined such as $m=L=10\omega=1$. Other parameters are $A=4$, $U_0=10/2\pi$, $\Gamma=10$. The directed current  vanishes in the overdamped limit (large frictions $\gamma$) for the shift-symmetric potential ($a=1$) because of  {\it hidden} symmetries. }
\label{fig:shiftsymm:shiftsymm} 
\end{figure}

A large class of Brownian motor systems, from particles in solution \cite{rou94} to vortices in superconductors \cite{silva2006} and atoms in dissipative optical lattices \cite{gommers2005}, corresponds to  a Brownian particle diffusing in a periodic potential under the action of a driving force with zero-average. The particle's motion is  described by the following Langevin equation
\begin{equation}
m\ddot{x}=-\gamma\dot{x}+\mathcal{F}(x,t)+\xi(t),
\label{eq:langevin}
\end{equation}
 where $\gamma$ is the friction coefficient, $\mathcal{F}(x,t)$ a generic deterministic force, and $\xi(t)$ a fluctuating force, modelled as a Gaussian white noise with autocorrelation $\langle \xi(t)\xi(t')\rangle=2\Gamma\delta(t-t')$, with the noise strength $\Gamma$ related to the temperature $T$ of the environment via the fluctuation-dissipation relation $\Gamma=\gamma k_B T$. The directed current is defined as $\langle v\rangle=\lim_{t\to\infty}\langle x(t)\rangle/t$, where the angle brackets denote average over noise realizations.  For finite noise strengths $\Gamma>0$, ergodicity implies $\langle v\rangle=\lim_{t\to\infty}x(t)/t$.
In very small systems, from the nanoscale to the microscale, the Brownian dynamics of small particles is frequently in the overdamped regime, where inertia effects---the term $m\ddot{x}$ in (\ref{eq:langevin})---can be neglected. This is the regime of interest here.

The standard symmetry analysis \cite{flach2000,reimann02,hanmar09,denisov14} relies on the identification of transformations which leave the equation  of motion  (\ref{eq:langevin}) unchanged and reverse the sign of the particle momentum. Trajectories with opposite momentum are equivalent, with a net null contribution to the directed current, which thus turns out to be zero. We will show that this picture does not fully capture the basic principles behind the operation of  Brownian ratchets. To do this, we consider a more general approach \cite{niurka2015}, and  regard the directed current $\langle v\rangle$ as a generic functional of the driving force $\mathcal{F}$, thus using the notation $v[\mathcal{F}(x,t)]$. Several properties follow from symmetry considerations. 

First, due to the vectorial nature of both the force $\mathcal{F}$ and the current, the transformation $x\to-x$ yields the following 
 property
\begin{equation}
v[-\mathcal{F}(-x,t)]=-v[\mathcal{F}(x,t)] .
\label{eq:spaceinv}
\end{equation}
Second, an arbitrary translation along the $x$ or $t$ axis does not alter the current, i.e.
\begin{equation}
v[\mathcal{F}(x,t)]=v[\mathcal{F}(x+x_0,t)]=v[\mathcal{F}(x,t+t_0)].
\label{eq:transl}
\end{equation}
Let us consider now a {\it forced ratchet}, i.e., $\mathcal{F}(x,t)=f(x)+F(t)$, where $f(x)=-\partial U(x)/\partial x$ is a conservative force and $F(t)$ a driving force.

If the system is spatially symmetric with respect a certain point $x_0$, then the potential satisfies $U(x+x_0)=U(-x+x_0)$. Without loss of generality, we choose the coordinate's origin such that $x_0=0$. Then $-f(-x)=f(x)$, which together with  (\ref{eq:spaceinv}) yields a characteristic property of spatially symmetric systems
\begin{equation}
v[f(x)-F(t)]=-v[f(x)+F(t)].
\label{eq:spasymm}
\end{equation}
A shift-symmetric force is defined as $F(t+t_0)=-F(t)$ ---for periodic drives $t_0=\tau/2$, where $\tau$ is the period, $F(t+\tau)=F(t)$. The direct application of properties  (\ref{eq:spasymm}) and (\ref{eq:transl}) yields no current for shift-symmetric forces in spatially symmetric systems,
\begin{eqnarray}
v[f(x)-F(t)]&=&v[f(x)+F(t+t_0)]=v[f(x)+F(t)]\nonumber\\
&=&-v[f(x)+F(t)]. \label{eq:symm:proof}
\end{eqnarray}
This is a well known result of spatially symmetric systems, already captured by the standard symmetry analysis \cite{flach2000,reimann02,hanmar09,denisov14}. However, our current approach reveals two additional symmetries for overdamped one-dimensional systems, which are not captured by the standard approach. They are:
\begin{eqnarray}
 v[f(-x)+F(t)]=v[f(x)+F(t)],  \label{eq:S1}\\
 v[f(x)+F(-t)]=v[f(x)+F(t)]. \label{eq:S2}
\end{eqnarray}
A proof of (\ref{eq:S1})--(\ref{eq:S2}) based on the Smoluchowski equation is given in the Supplemental Material \cite{suppl}. \nocite{risken1984,fau95,grier05,arz11,jones13,lee1999,shalom2005,wambaugh1999,nori2006,jones13,cecile99,schiavoni2003,gommers2006,
cubero2010,wichol12}
The symmetries (\ref{eq:S1}) and (\ref{eq:spaceinv}), together with (\ref{eq:transl}), yield the following property for shift-symmetric potentials,
\begin{eqnarray}
&v[f(x)+F(t)]=v[f(-x)+F(t)]=-v[-f(x)-F(t)]\nonumber\\
&=-v[f(x+L/2)-F(t)]=-v[f(x)-F(t)].
\end{eqnarray}
This is the same property as Eq. (\ref{eq:spasymm}), and proceeding as before it implies current suppression when combined with a shift-symmetric driving force. 
Therefore, quite counter-intuitively, in one-dimensional overdamped systems, the condition for current suppression for systems with shift-symmetric potentials---like the one shown in Fig.~\ref {fig:shiftsymm:shiftsymm}(b)---is, despite being spatially asymmetric,  the same as for spatially symmetric potentials. Figure~\ref {fig:shiftsymm:shiftsymm}(c) confirms this unexpected behavior for large enough frictions. In the underdamped regime this property is not satisfied exactly. Nevertheless, even in this regime the overdamped symmetry (\ref{eq:S1}) identified here has a lasting effect: The zero-current point determined by the overdamped symmetry is displaced to a lower value of the symmetry parameter $a$ in the underdamped regime. Thus, the overdamped symmetry (\ref{eq:S1}) determines a current reversal in the underdamped regime.

The discovery of hidden symmetries reported above does not represent the only departure from the conclusions which can be drawn from the standard symmetry analysis.
The presence of instabilities may also alter the picture, as trajectories which are solutions of the equations of motion with opposite momenta may have very different stability properties, and thus result into a total non-zero contribution to the system current. Such a scenario of spontaneous symmetry breaking is best illustrated via a specific case study.

\begin{figure}
\includegraphics[width=8cm]{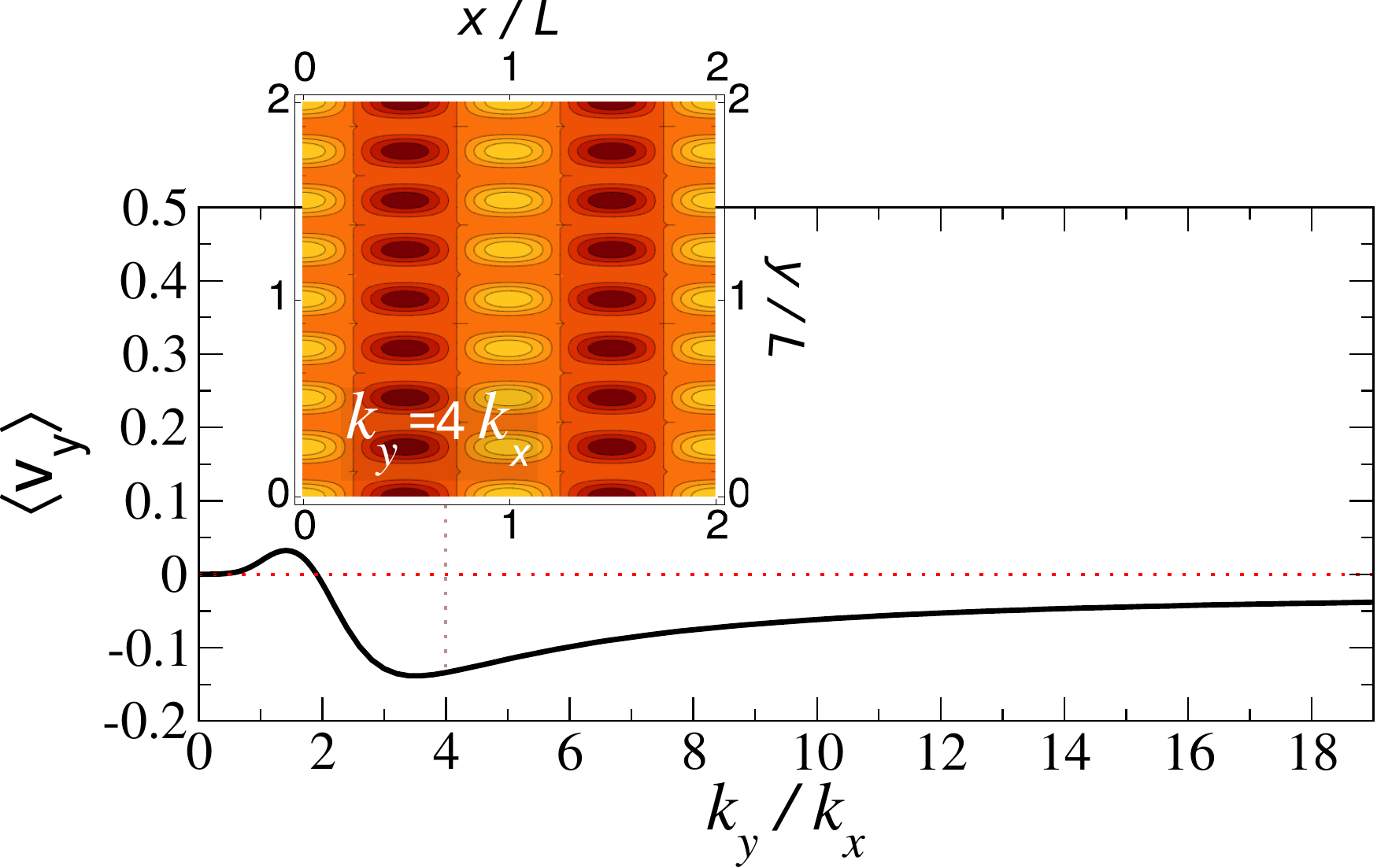}
\caption{Breaking of symmetries  (\ref{eq:S1})--(\ref{eq:S2}) in a 2D overdamped system. The driving force is ${\bf F}(t)=A[\cos(\omega t){\bf e}_x+\cos(2\omega t+\pi/2){\bf e}_y]$, i.e. a bi-harmonic drive split \cite{denisov2008} in two perpendicular directions. The potential is $U(x,y)=U_0\cos(k_x x)[1+\cos(k_y y)]$, which  is spatially symmetric in both directions, and shift-symmetric along the $x$-direction.  The current is produced in the $y$-direction only---due to the symmetry (\ref{eq:symm:proof}) in the $x$-direction---through  the coupling with the dynamics in the $x$-direction. Despite the driving force being anti-symmetric, a nonzero current is observed when $k_x$ and $k_y$ are comparable.
Reduced units are defined such as $m=k_x=\omega=1$.  Other parameters $U_0=\gamma=50$, $A=2\gamma$, $\Gamma=0.1\gamma^2$. The inset illustrates the potential landscape for $k_y=4k_x$, with $L=2\pi/k_x$.}
\label{fig:2D} 
\end{figure}

\begin{figure}
\includegraphics[width=8cm]{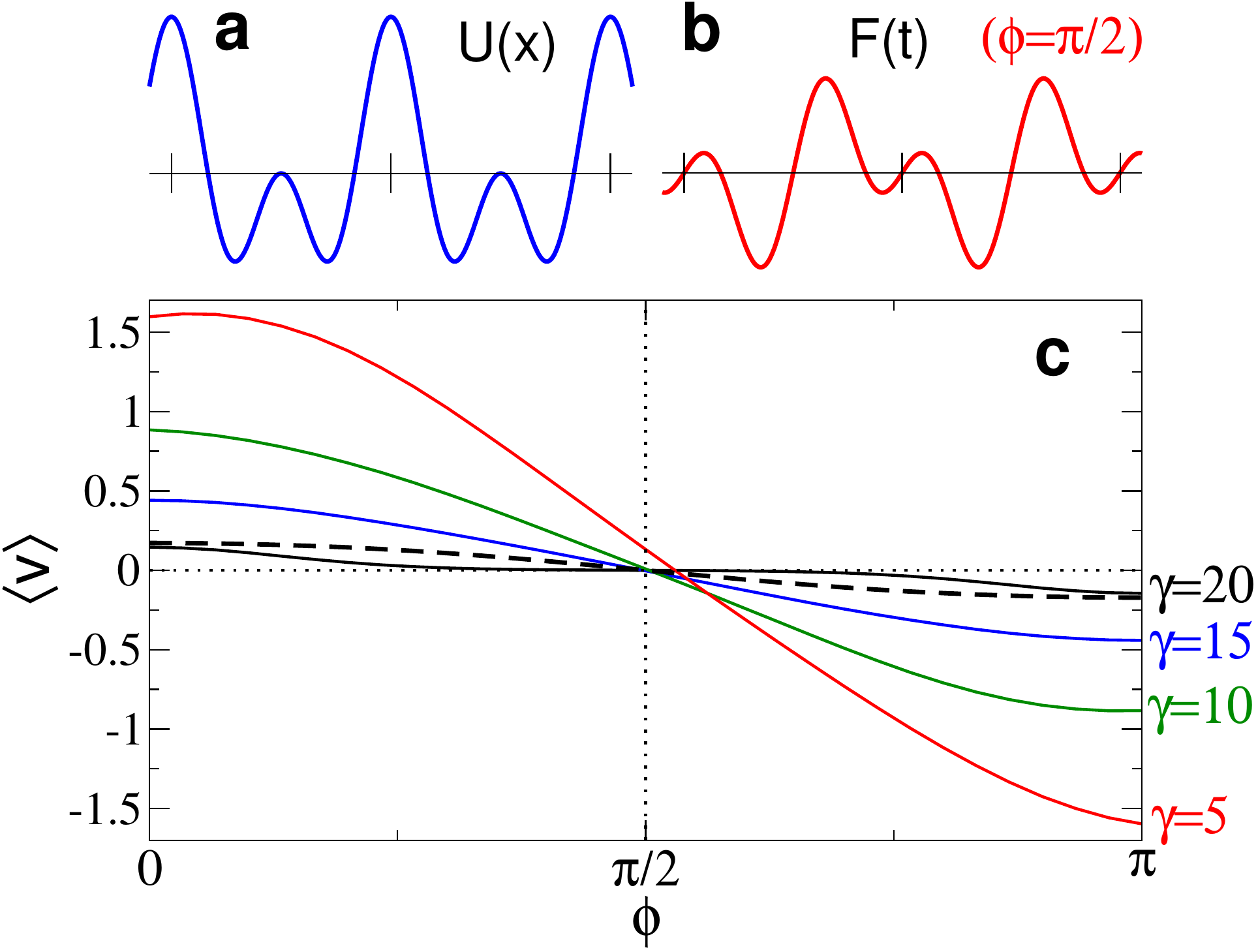}
\caption{Current suppression in one-dimensional overdamped systems with a spatially symmetric potential and applied anti-symmetric forces. The driving force has a biharmonic shape, $F(t)=A[\cos(\omega t)+\cos(2\omega t+\phi)]$. (a) Spatially symmetric potential $U(x)=U_0[\cos(kx)+\cos(2kx)]$. (b) The driving force $F(t)$ is anti-symmetric when $\phi=\pi/2$. (c) Directed current as a function of the driving phase $\phi$, for different levels of damping. Reduced units are defined such as $m=k=\omega=1$. Other parameters are $U_0=20$, $A=\Gamma=40$. The dashed line shows, for comparison, the current $(\langle v_y\rangle)$ for the same driving force applied in the $y$-direction and a two-dimensional potential $U(x,y)=U_0\cos(k x)[1+\cos(4k y)]$ in the overdamped regime ($\gamma=U_0=50$, $A=2\gamma$, $\Gamma=0.1\gamma^2$). 
}
\label{fig:sympot:antisym} 
\end{figure}

In the overdamped regime, from every solution $x(t)$, the trajectory $\widetilde{x}(t)=x(-t)+L/2$ is also a solution of (\ref{eq:langevin}) provided the potential is shift-symmetric. It corresponds to a transformed random force $\widetilde{\xi}(t)=-\xi(-t)$, which is statistically equivalent to $\xi(t)$, and a driving force $\widetilde{F}(t)=-F(-t)$. Following the standard symmetry analysis, no current is expected when anti-symmetric driving forces $F(t+t_0)=-F(-t+t_0)$ are applied \cite{flach2000,reimann2001,hanmar09,denisov14}---an appropriate  choice of the time origin yields $t_0=0$.  This prediction is correct in one-dimensional systems, as readily verified by numerical simulations. However the same reasoning predicts no current also in the case of higher dimensions, a result  which is contradicted by our numerical simulations, as shown in Fig.~\ref{fig:2D} for a two-dimensional potential and an applied split biharmonic drive, as well as by independent results by Peter Reimann's group (see Ref. \cite{speer}, page 16). The presence of instabilities is the key to understand such an unexpected, spontaneous symmetry-breaking behaviour. The standard analysis fails to account for the actual instability of the transformed solutions $\widetilde{x}(t)$, which makes them very unlikely. Even in the noiseless limit, the above transformation maps stable oscillations about the potential minima into highly unstable oscillations about potential maxima \cite{denisov02}.  We have verified via numerical simulations that, given a stable solution $x(t)$, the transformed solution $\tilde{x}(t)$ is unstable and thus quickly collapses onto $x(t)$. This occurs both in one dimension, and in higher dimensional systems  \cite{suppl}. 
 Given that instabilities destroy the mechanisms of current suppression due to the contributions of a trajectory and the transformed one, 
the observed suppression of directed transport in one-dimensional systems  must be associated to a different mechanism.
This suppression under anti-symmetric forces is actually a consequence of  the symmetry  (\ref{eq:S2}), which yields no current for systems---which include spatially symmetric as well as spatially shift-summetric systems of interest here---satisfying the property (\ref{eq:spasymm}):
\begin{eqnarray}
v[f(x)+F(t)]&=&-v[f(x)-F(t)]=-v[f(x)+F(-t)]\nonumber\\
&=&-v[f(x)+F(t)].
\end{eqnarray}
A consequence of this analysis is that also truly spatially symmetric systems should exhibit no current in one-dimensional overdamped systems when anti-symmetric forces are driving the system. This phenomenon is illustrated in Fig.~\ref{fig:sympot:antisym}.  It was already experimentally observed in Ref.  \cite{nori07}, but it remained unexplained until  the present work.  These results are a confirmation of the validity of the approach based on a more general symmetry analysis, which does not rely on the direct analysis of the solutions of the equation of motion.

\begin{figure}
\vspace{1em}
\includegraphics[width=8cm]{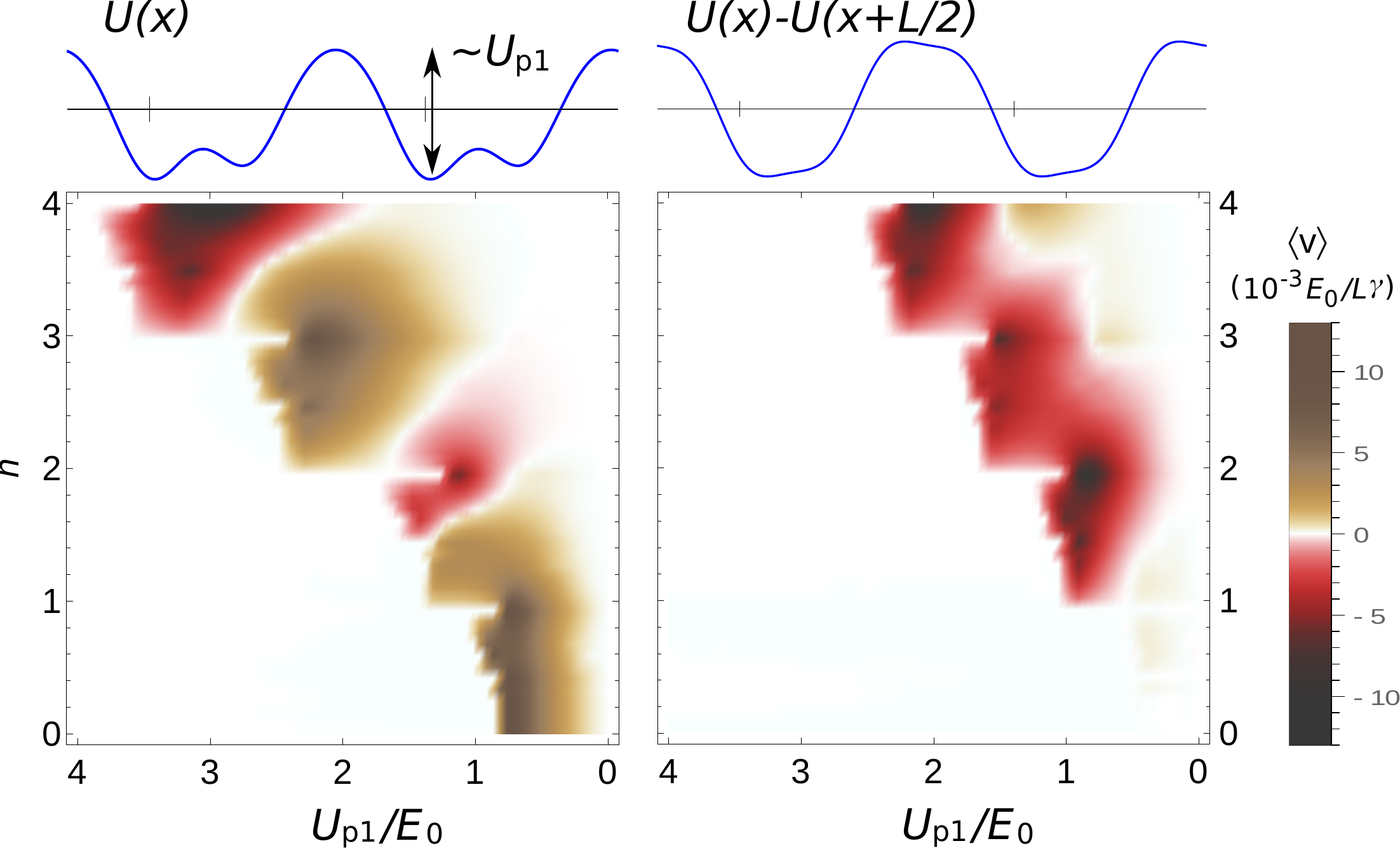}
\caption{Cancellation of transport, via the use of shift-symmetric potentials, for the one-dimensional overdamped system of interacting particles of Ref.~\cite{silva2006}. The bottom panels show the net chain current as a function of the number of particles per period, $n$, and the potential depth $U_{p1}/E_0$, with the left panels referring to the original one-particle potential $U(x)$ (depicted on the upper panel), and the right panels  to  a shift-symmetric potential built from the former  as $U_\mathrm{ss}(x)=U(x)-U(x+L/2)$. The interaction between the particles is accounted for by the pair potential $V_\mathrm{int}(r)=-E_0\ln(r)$, with $r$ the particle separation. The system is driven by a single-harmonic force acting on each particle, which is both shift-symmetric and anti-symmetric. The parameters are the same as in Fig.~2 of \cite{silva2006}. }
\label{fig:Silva} 
\end{figure}

It is worth stressing that in the present discussion the dimensionality of the system corresponds to the number of spatial degrees of freedom taking part into the rectification mechanism, and not necssarily to the dimensionality of the potential landscape.
The violation of the symmetries  (\ref{eq:S1})--(\ref{eq:S2}) in the above 2D overdamped setup is due to a rectification mechanism taking place in the two perpendicular directions. However, the symmetries (\ref{eq:S1})--(\ref{eq:S2}) are not restricted to strictly one-dimensional systems, they are still present in higher-dimensions overdamped systems provided the rectification mechanism involves one spatial dimension only. For example, the dashed line in Fig.~\ref{fig:sympot:antisym}(c) shows the suppression of current for anti-symmetric driving for the same 2D system shown in Fig.~\ref{fig:2D} when the bi-harmonic driving force is applied in the $y$-direction only. Additional examples are shown in \cite{suppl}.
\paragraph*{Discussion ---}
The hidden symmetries identified in the present work are of relevance to current experiments, as well as they allow to recast known results within a more general theoretical framework. This is well exemplified by two specific case studies presented in the following.

The first case study correspond to  the system of a.c.-driven vortices trapped in a superconductor experimentally studied in Ref.~\cite{silva2006}. Here, inter-particle interactions provides an additional  path to escape from the symmetries (\ref{eq:S1})--(\ref{eq:S2}).  Our results of Fig.~\ref{fig:Silva} precisely refer to the one-dimensional system of interacting Brownian particles, successfully used in Ref.~\cite{silva2006} to explain the multiple current reversals observed on a.c.-driven vortices trapped in a superconductor. Despite not being strictly satisfied, the influence of the symmetries  (\ref{eq:S1})--(\ref{eq:S2}) is quite noticeable, cancelling the ratchet effect and most of the current reversals in regions of the parameter space where the appearance of a current is not directly related to particle interactions. 

As second case study, we refer to the celebrated {\it flashing ratchet model} \cite{ajd92,rou94,bader99,chou99,vanou99,parrondo99}, where the ratchet potential is periodically switched on and off in the absence of any additional additive driving $F(t)$--- i.e., here $\mathcal{F}(x,t)=-\partial U(x,t)/\partial x$---and more specifically to the known \cite{kanada04} result that a flashing shift-symmetric potential can not produce directed motion. The theoretical framework, and the related new symmetries, introduced here allows for a simple explanation of such a result.

Generally, in one-dimensional overdamped systems the following symmetry is satisfied \cite{suppl}
\begin{equation}
v[\mathcal{F}(-x,-t)]=v[\mathcal{F}(x,t)].\label{eq:flash}
\end{equation}
In two-state systems that are periodically switched, reversing the direction of time has no effect, $v[\mathcal{F}(x,-t)]=v[\mathcal{F}(x,t)]$, a fact which together with (\ref{eq:spaceinv}), (\ref{eq:flash}) and (\ref{eq:transl}) yield no current for shift-symmetric potentials,
\begin{eqnarray}
v[\mathcal{F}(x,t)]=-v[-\mathcal{F}(-x,t)]=-v[-\mathcal{F}(x,-t)]= \quad\nonumber \\
-v[-\mathcal{F}(x,t)]=-v[-\mathcal{F}(x+L/2,t)]=-v[\mathcal{F}(x,t)].\quad
\end{eqnarray}
Therefore, a flashing ratchet with a shift-symmetric potential, regardless if it is spatially asymmetric or not, can not  produce directed motion, thus showing that  in overdamped systems  shift-symmetric potentials behave like spatially symmetric ones.

\paragraph*{Conclusions ---}
The present work addresses the outstanding issue of providing a general theoretical framework for the identification of symmetries not captured by the standard symmetry analysis, examples of which were already 
given in previous works \cite{denisov02,kanada04} with ad-hoc treatments.
We have proven the existence in a prototypical 1D overdamped system of {\it  hidden} symmetries, which escape the identification by the standard symmetry analysis and require different theoretical tools for their revelation. Though rigorously not satisfied in higher-dimensional systems, their effects are shown to be still noticeable in them. Our results pave the way to new mechanisms of manipulating transport. The hidden symmetries determine in fact current reversals, which can be used to precisely control transport and implement mechanisms for particles separation.  Specific realizations for optical tweezers and cold atom set-ups are discussed in the Supplemental Material \cite{suppl}.

\begin{acknowledgments}
 Financial support from the Royal Society (IE130734) (DC and FR), and the Leverhulme Trust (Grant RPG 2012 – 809) (FR) is acknowledged. 
\end{acknowledgments}

 

%


\end{document}